\documentstyle[11pt,dunk2001_asp,twoside, psfig, epsf]{article}
\def\edcomment#1{\iffalse\marginpar{\raggedright\sl#1\/}\else\relax\fi}
\reversemarginpar

\begin{document}
\title{Penetrating the Mask: The Gravitational Torque of Bars}
 \author{D. L. Block}
\affil{University of the
Witwatersrand,
Johannesburg, South Africa}
\author{R. Buta}
\affil{University of Alabama, Tuscaloosa, Alabama, USA}
\author{I. Puerari}
\affil{INAOE,
Tonantzintla, Puebla, M\'exico}
\author{J. H. Knapen}
\affil{Isaac Newton Group of Telescopes, La Palma, Spain}
\author{B. G. Elmegreen}
\affil{IBM Research Division, Yorktown Heights, NY, USA}
\author{S. Stedman}
\affil{Univ of Hertfordshire, Hatfield, Herts, UK}
\author{D. M. Elmegreen}
\affil{Vassar College, Dept Phys \& Astronomy, Poughkeepsie, NY, USA}

\begin{abstract}
The Hubble classification scheme of galaxies is based on blue-light
appearance. 
Atlases reveal the rich variety of responses of the Population I component
(`the mask') of gas and dust to the underlying, older, stellar population.
However, the Population I component may only constitute 5 percent of the
dynamical mass of the galaxy; furthermore, dusty masks are highly effective
in hiding bars. In the 1960s, Ken Freeman presented a meticulous study
of the dynamics of bars at a time when nonbarred galaxies were called
``normal'' spirals and barred galaxies were regarded as curiosities.
Now we know that it is more ``normal'' for a galaxy to be barred than
to be nonbarred. What is the range for the gravitational torques of bars?
We describe here a recently developed method for deriving relative bar torques
by using gravitational potentials inferred from near-infrared light 
distributions. We incorporate a bar torque class into the Block/Puerari
dust-penetrated galaxy classification system.
We find a huge overlap in relative bar torque between Hubble (Sa, Sb,
...) and (SBa, SBb, ...) classifications.
Application of the method to the high redshift universe is
briefly discussed.
\end{abstract}

\section{Introduction}

Bars have been recognized in galaxies since the time of Curtis
(1918) and Hubble (1926), although the bar-like structure of the LMC
was already beautifully portrayed in a remarkable sketch by J. Herschel
in 1847 (Figure 1). Bars are among the most interesting
features of galaxies that present a clear challenge to theorists. 
It is not surprising, then, that bars
attracted the early attention of Ken Freeman, who applied his mathematical
expertise to understanding them in a remarkable series of papers
published in the 1960s (Freeman 1965; 1966a,b,c). Of course, at
that early stage, bars were regarded mostly as dynamical curiosities rather
than representing a major topic of research, which still continues to unfold
at the present time. Nevertheless, Ken pursued them initially in his
Cambridge dissertation work and later developed a model of asymmetric
barred galaxies in a collaborative effort with G. and A. de Vaucouleurs
(de Vaucouleurs, de Vaucouleurs, and Freeman 1968). 

Ken's work, in a way,
culminated with his fine review of barred galaxies with Gerard de
Vaucouleurs, the first really significant review of the subject
(de Vaucouleurs and Freeman 1972). This long article focussed mainly
on the features of
late-type, asymmetric barred spirals of the Magellanic type, and
also outlines an early interpretation of the nature of the inner rings
of SB(r)-type galaxies in terms of a class of resonant periodic
orbits. It is interesting to note that at the time of 
Ken's work, the status of rotation curves suggested that bars
rotated as rigid bodies (see also Figure 2).     

It is also remarkable that only recently have people perceived just how strong bars are in their host galaxies, that is,
how significant is their forcing relative to the dominant mass
components. This was previously known only in a few individual cases that
had been the subject of detailed dynamical models (e.g., Lindblad, Lindblad,
and Athanassoula 1996). For the general
galaxy population, the quantification of bar strength in terms of
forcing had to wait for the advent of routine near-infrared
imaging. We have been engaged in dust-penetrated galaxy classification
for some time now, and describe here a recently-developed method for
quantifying the gravitational torques of bars using potentials
inferred from near-IR images. Bar strength is important in galaxy
morphological studies because phenomena such as gas inflow, angular
momentum transfer, noncircular motions, lack of abundance gradients, nuclear
activity, starbursts, and the shapes and morphologies of rings and spirals, 
may all be tied in various ways to
the effectiveness with which a bar potential influences the motions
of stars and gas in a galactic disk (e.g., Sellwood \& Wilkinson
1993; Buta \& Combes 1996; Knapen 1999).

\begin{figure}
\plotone{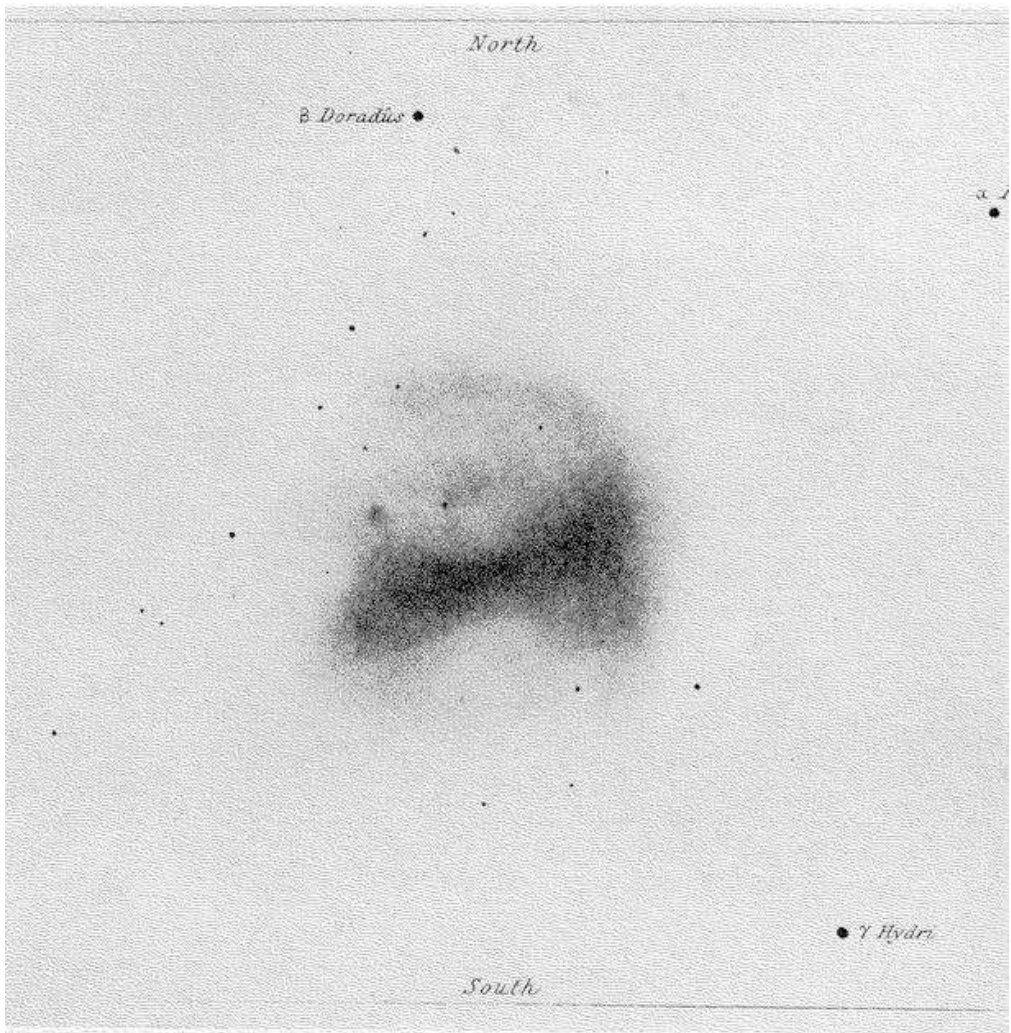}
\caption{The bar in the Large Magellanic Cloud is beautifully portrayed in this naked
  eye drawing by Sir John Herschel in 1847. Reproduced in de
  Vaucouleurs 
\& Freeman (1972).}
\end{figure}

\section{Removing the Mask: Dust-Penetrated Galaxy Classification}

Two hours north of Dunk Island, our meeting locale, lies the land of 
Papua New Guinea. It is the country
where masks still dance. Customs have remained unchanged for centuries.
Where men and women, as if from the stone age, meet the New Millennium.
In Roget's Thesaurus, we find the following:

{\bf Mask:} [noun] screen, cloak, shroud. [verb] to camouflage, to
make opaque, to disguise.

Optically thick dusty domains in galactic disks can completely
camouflage or disguise
underlying stellar structures. {\it Cosmic dust grains act as masks}.
The dust masks obscure whether or not the dust lies in an actual screen
or is well intermixed with the stars. The presence of dust and the
morphology of a
galaxy are inextricably intertwined: indeed, the morphology of a galaxy can
completely change once the Population I disks of galaxies -- the
masks -- are dust penetrated (e.g., Block
and Wainscoat 1991; Block et al.,
1994, 2000).

The classification of galaxies has traditionally been inferred from
photographs or CCD imaging shortward of the 1$\mu m$ window, where
stellar Population II disks are not yet dust-penetrated. Images
through an $I$ (0.8 $\mu m$) filter can still suffer from
attenuations by dust at the 50\% level. The
NICMOS and
other near-infrared camera arrays offer unparalleled opportunities for
deconvolving the Population I and II morphologies, because the opacity at
$K$ -- be it due to absorption or scattering
-- is always low. The extinction (absorption+scattering) optical depth at $K$
is only 10\% of that in the V-band.

Many years before the advent of large format near-infrared camera
arrays, it
became increasingly obvious from rotation curve analyses that
optical Hubble type is not correlated with the evolved Population II
morphology. This was already evident in the pioneering work of Zwicky
(1957) when he published his famous photographs  
showing the `smooth red arms' in M51.
In the {\it Hubble Atlas} and other atlases showing optical
images of galaxies, we are
looking at masks: at the gas, not the stars,
to which the properties of rotation
curves are inextricably tied.

\section{A duality in spiral structure}

There is a fundamental limit
in predicting what an evolved stellar disk might
look like (Block et al. 1994, 2000). The greater the degree of decoupling,
the greater is the uncertainty.
The fact that a spiral might be
flocculent in the optical
is very important, but it is equally important to know whether or not
driving the dynamics is a  
grand design old stellar disk.

Decouplings between stellar and gaseous disks are cited in many
studies including Grosb{\o}l \& Patsis (1998), Elmegreen et al.
(1999),  Block et al. (2000) and Puerari et al. (2000). The Hubble type
of a galaxy does not dictate its dynamical mass distribution (Burstein
\& Rubin 1985).  This is confirmed by examining  
Fourier spectra, for example, of the evolved disks of NGC
  309 (Sc) and NGC 718 (Sa); these spectra  are almost identical
  (Figs. 11 and 12 in Block et al. 2001a). 

\begin{figure}
\plotone{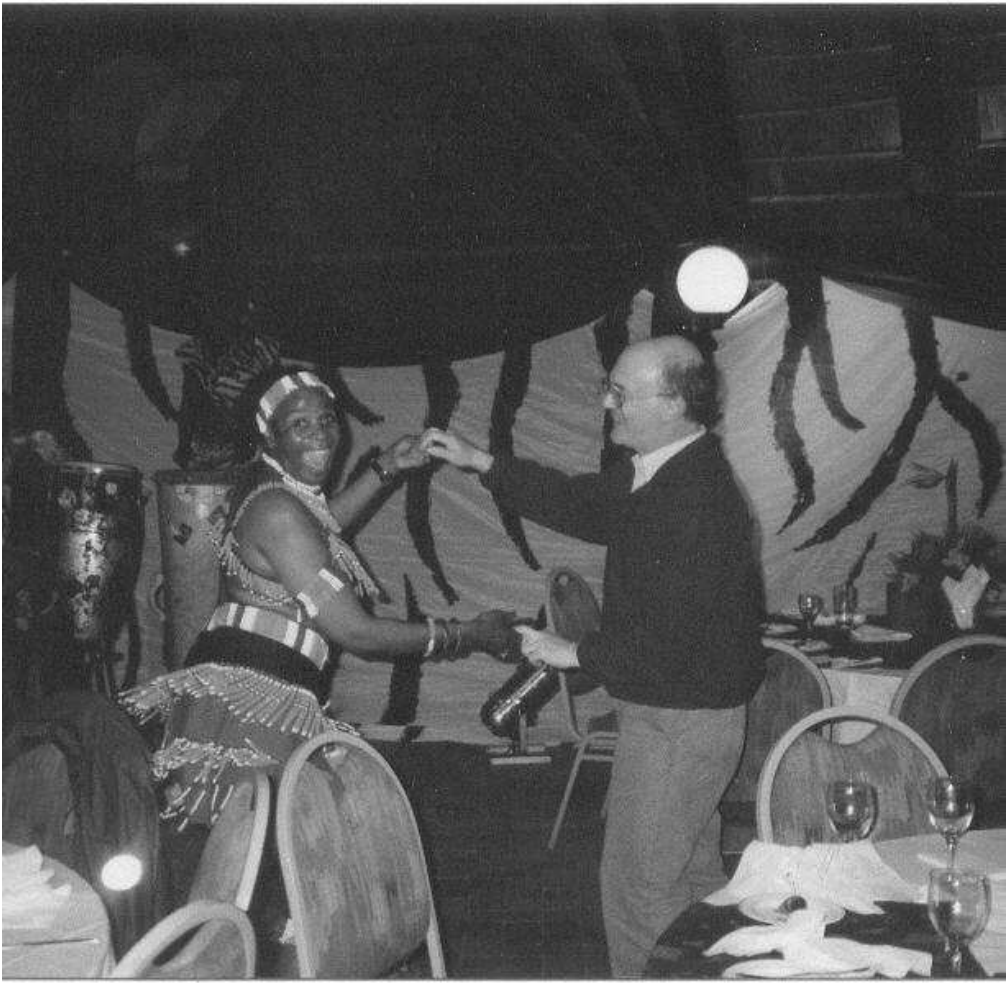}
\caption{Like the human frame, bars do not behave as a rigid body. To
  demonstrate this point clearly is Ken Freeman, seen here with body
  parts moving differentially to
  the thundering rhythm of drums in Africa. Photo by J. Mayo Greenberg,
  reproduced from the Kluwer
  volume {\it `Toward a New Millennium in Galaxy Morphology'}
  (eds. D.L. Block, I. Puerari, A. Stockton and D. Ferreira, Dordrecht
  2000).}
\end{figure}

\section{Bar Strength: the 60s versus the 00s}

In the 60s, bar strength was something that was judged visually on blue-light
photographs.  Galaxies were recognized as either S or SB in Hubble's (1926,
1936) system, or as SA, SAB, SB in de Vaucouleurs' (1959) revised Hubble
system. It was assumed that an SB galaxy has a stronger bar than an
S galaxy, and that the sequence SA-SAB-SB was a sequence of increasing
average bar strength 
However, neither the Hubble nor the de Vaucouleurs bar
classifications
can be expected to be accurate measures of bar strength in individual
cases because
apparent bar strength is impacted by wavelength, the effects of
extinction
and star formation, inclination and bar orientation
relative to the line of sight, and also on observer interpretations.
The percentage of unbarred galaxies in the {\it Carnegie Atlas}
dramatically drops from 70\% to only 27\% when Sa, Sb, Sc spirals are
mask-penetrated (Eskridge et al. 2000; see also Knapen, 
Shlosman \& Peletier 2000, Block \& Wainscoat 1991).
Thus, the near-IR is the best wavelength regime for
quantifying the strength of bars. A simple, easily reproducible quantitative 
near-infrared morphological
classification scheme for spirals that accounts for bar strength,
dominant harmonic, and arm pitch angle class has been proposed by Block \&
Puerari (1999) and Buta \& Block (2001) and is seen in Figure 3.  

\begin{figure}
\plotone{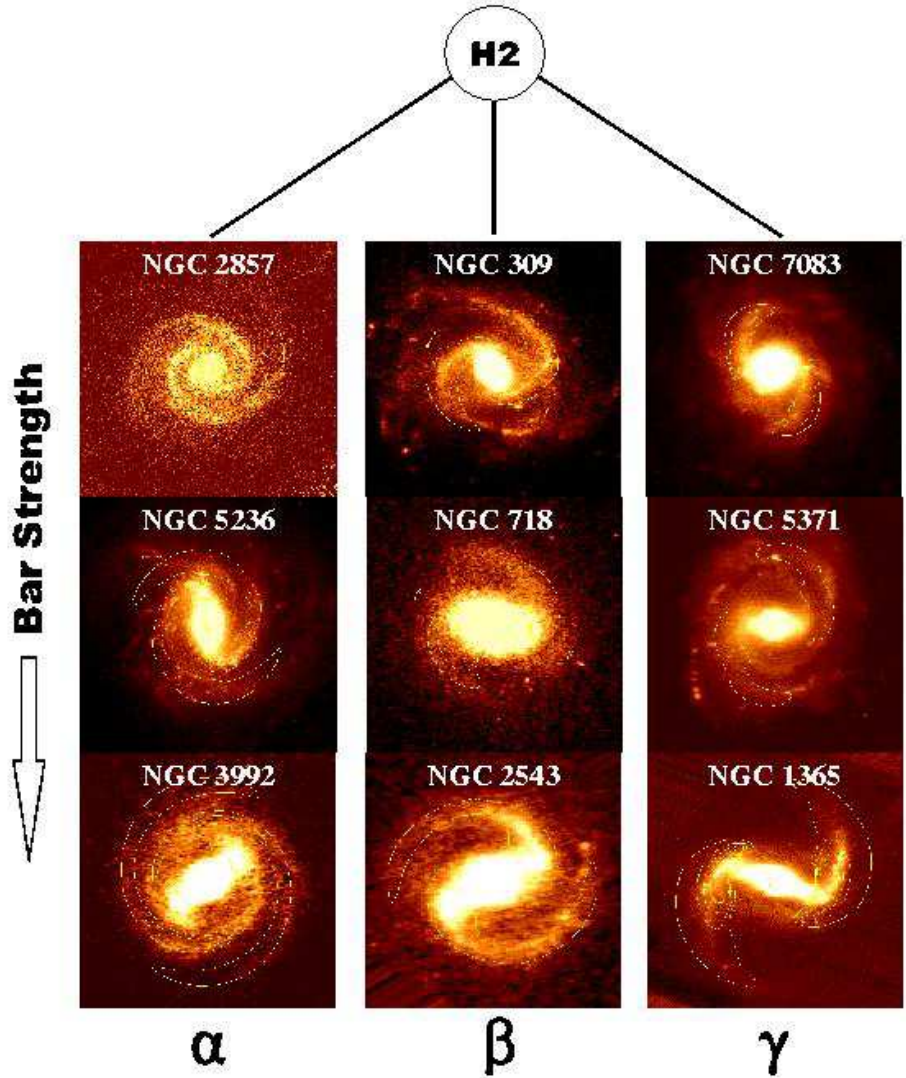}
\caption{Spiral galaxies in the dust penetrated regime are binned
according
to three quantitative criteria: H$m$, where $m$ is the dominant
Fourier harmonic (illustrated here are the two-armed H2 family); 
the pitch angle families $\alpha$, $\beta$ or $\gamma$
and thirdly the bar
strength,
derived from the gravitational torque (not ellipticity) of the
bar. Early type b spirals (NGC 3992, NGC 2543, NGC 7083, NGC 5371
and NGC 1365) are distributed within all three families ($\alpha$,
$\beta$ and $\gamma$). Hubble type and dust penetrated class are 
uncorrelated.}
\end{figure}

\section{Bar Strengths Derived from Gravitational Torques}

Any reliable measure of bar strength ought to involve forces, because
this is how the influence of a bar is actually felt. Unlike the axisymmetric
background, a bar involves both radial and tangential forces. 
The most elegant definition of bar strength in terms of these force components
was proposed by Combes and Sanders (1981). Given the gravitational
potential $\Phi(R,\theta)$ in the disk plane,
these authors defined the bar strength at radius $R$ as

\begin{equation}
Q_T(R) = {F_T^{max}(R) \over F_0(R)} = 
{{{1\over R}\bigl{(}{\partial \Phi(R,\theta)\over \partial\theta}\bigr{)}_{max}} \over {d\Phi_0(R)\over dR}} 
\end{equation}

\noindent
where $F_T^{max}(R)$ 
represents the maximum amplitude of the tangential force
and $F_{0}(R)$ is the mean axisymmetric radial force, inferred
from the $m$=0 component of the gravitational potential.
In this approach, the strength of the bar is measured relative
to the axisymmetric forces of the background disk. A relatively
thin bar imbedded in a massive axisymmetric disk may have significant
tangential forces, but relative to the background radial force field,
it could be weak. Thus, measuring bar strength as a force ratio (rather
than an isophotal axis ratio) is an idea whose time has come.
Our gravitational potentials
are derived from near-IR images under the assumptions of a
constant mass-to-light ratio and an exponential vertical
scale height (Quillen, Frogel, \& Gonz\'alez 1994).

\begin{figure}
\plotone{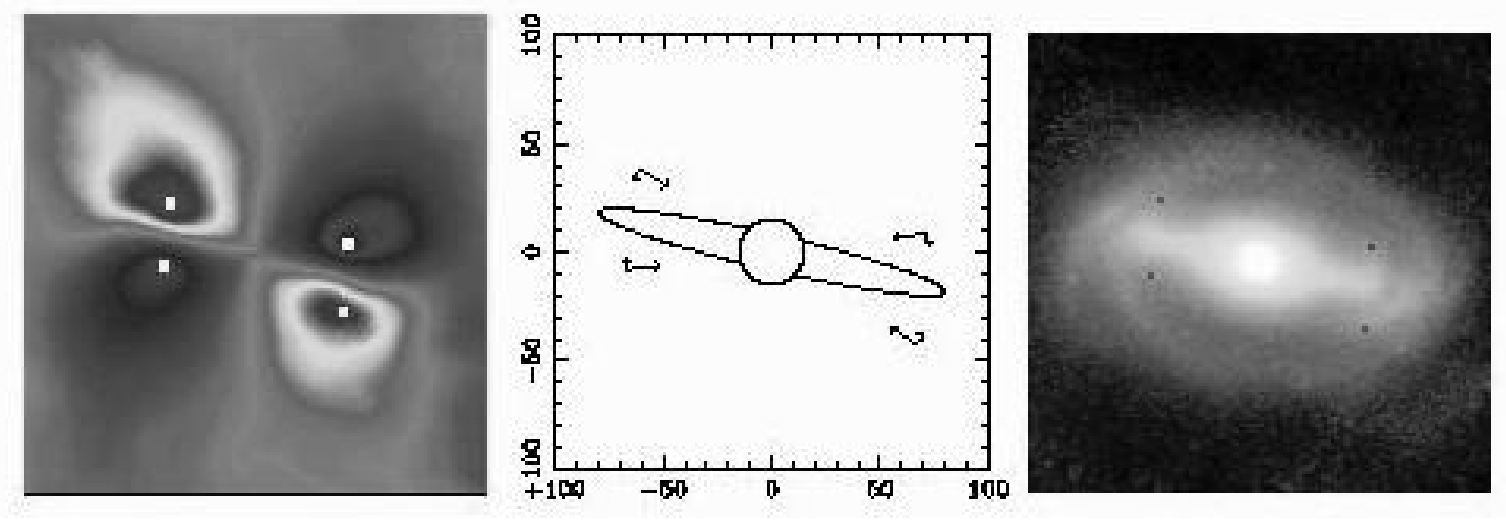}
\caption{The characteristic signature of a bar is this ratio map, or  
butterfly pattern. The map shows four well-defined regions where the
  tangential-to-radial force ratio reaches a maximum or minimum around
  or near the ends of the bar. Seen at left is the ratio map for NGC
  1433; the galaxy appears at right. The sign of the tangential force
  changes from quadrant to quadrant (central schematic) and absolute
  values are considered in our gravitational bar torque method.}
\end{figure}

A tangential-to-radial force ratio map reveals a butterfly pattern
that is the characteristic signature of a bar (see Figure 4, adapted from
Buta \& Block 2000).
The actual force ratio changes sign from quadrant to quadrant
relative to the bar axis, because the total force in the plane
is slightly offset towards the ends of the bar.
In the case of
perfect symmetry, $|Q_T|$ would reach a maximum at the same
radius and angle relative to the bar in each quadrant. However,
slight asymmetries and/or noise can make these maxima different 
in each quadrant. If we let $Q_{bi}$ = $|Q_T|^{max}$
in quadrant $i$, then the average of these
four values can provide a single measure of bar strength for
a whole galaxy, if the gravitational potential is known. 
We call this average the {\it relative bar torque parameter}, $Q_b$.
The $Q_{bi}$ are known as the ``maximum points.''

Bar torque classes are defined in terms of intervals of $Q_b$. 
Bar class B$_c$=1 includes galaxies having relative torques
$Q_b$= 0.1$\pm$0.05 (meaning the tangential force reaches a
maximum of 10\% of the axisymmetric background radial force);
class 2 involves those with $Q_b$= 0.2$\pm$0.05,
etc., up to class 6 (Buta \& Block 2001).

Uncertainties in the bar torque
method are discussed in Buta \& Block (2001).
These include simple deprojection uncertainties, such as bulge
``deprojection stretch'', as well as
more complex uncertainties, such as variations
in the stellar mass-to-light ratio with position, the impact
of dark matter and bulge thickness, vertical resonances and
the thickening of bars, and the generally unknown 
vertical scale heights of disks. Future studies will involve
evaluating these uncertainties further and refining the method.

\begin{figure}
\plotfiddle{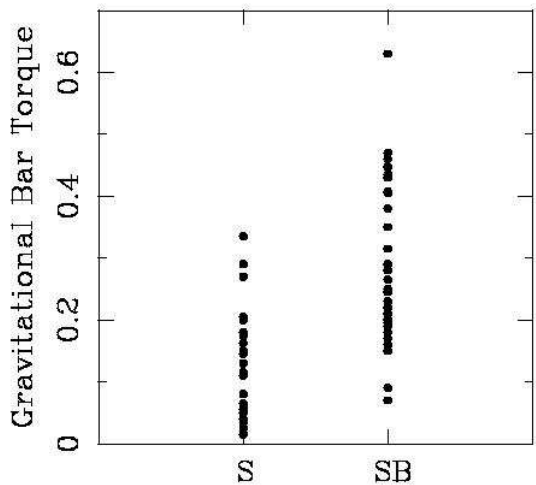}{6.5cm}{0} {50} {50} {-160} {-30}
\caption{ The threshold for calling a galaxy ``SB'' is bar
torque class 1 (where Q$_{b}$ ranges from 0.05 to 0.149). NGC
5371 (Sb/SBb) is an example of bar torque 1. The
Hubble classification does not make any further
discrimination on bar strength Q$_{b}$ beyond this 
threshold. We find that the bars with the strongest gravitational
torques reach a bar class of 6, where the maximum tangential force
reaches about 60\% of the mean radial force. Further details
in Block et al. (2001b).}
\end{figure}

\begin{figure}
\plotone{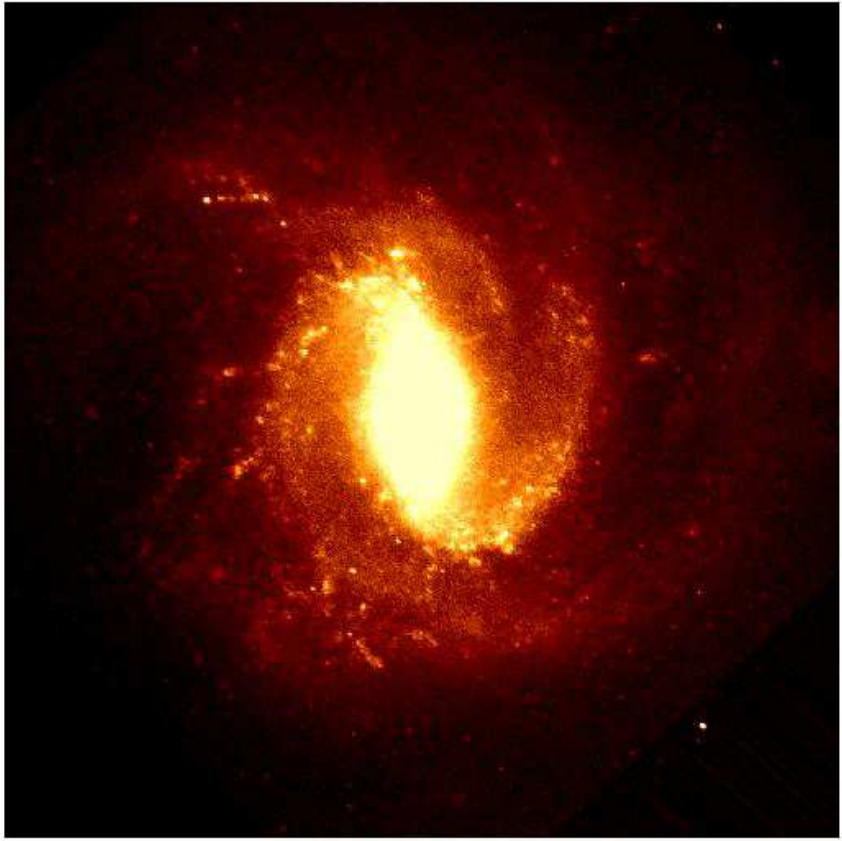}
\caption{A near-infrared image of NGC 5236 (M83) courtesy O.K. Park
  and K. Freeman, shows a highly
  elongated bar. Martin (1995) assigns the bar of M83 to his strong
  bar ellipticity class 7. The gravitational bar torque of M83 is,
  however, weak \
(bar
  torque class 2). The tangential forces in the environs of the bar in
  M83 only
  reach 
a maximum of  $\sim$20\% of the
  background mean axisymmetric
  radial force, whereas for galaxies with truly strong bar torques, 
these may reach maxima of 60\%. }
\end{figure}

\section{Hubble Classifications and Bar Torques}

From our studies (Buta \& Block 2001; Block et al. 2001b), we
have estimates of $Q_b$ for 75 galaxies at this time.
Figure 5 shows how this parameter correlates with optical
bar classification in the Revised Shapley-Ames Catalogue
(RSA, Sandage and Tammann 1981). A wide range of relative
bar torques characterizes each of the S and SB categories.
Category S (i.e., optically unbarred) in Fig. 5 includes
galaxies ranging from bar torque class 0 (e.g., NGC 628)
to bar class 3 (e. g., NGC 1042).  NGC 4321, a Hubble Sc
prototype,
has a bar class of 2. Likewise, NGC 4450 (Sab) is of bar class
2.

 Similarly, Hubble category SB in the RSA has a wide range of
bar strengths. This category
commences at
a bar torque of class 1 (e.g., NGC 5371) and
reaches bar class 5 (e.g., NGC 7741) and class
6 (e.g., NGC 7479).
In other words, the bar strengths of some
RSA SB galaxies may be {\em weaker}
than those found in
RSA unbarred spirals such as NGC 1042 (Sc; near-infrared
bar class 3). This is not due to the uncertainties in the
$Q_b$ method, but instead reflects the difficulties of making
reliable bar strength judgments in the visual Hubble system. The
work of Knapen et al. (2000) reaches this
identical
conclusion, using their independent definition of bar strength.

Martin (1995) classified bars according to {\it shape of isophotes}. One of
the strongest classes in Martin's sample is bar ellipticity class 7.
It is interesting to note that highly elliptical bars as measured by
Martin (1995)  
need {\it not} have strong gravitational bar torques (see Figure 6).

\section{Toward the future: Choice of instruments for NGST and Dust penetrated morphology in the
  high-redshift 
universe}


HDF morphology preferentially samples restframe UV light. Does the
 duality of spiral structure, found in our local Universe (section 2)
 persist at higher $z$? In order to
 explore the capability of NGST for undertaking stellar (as opposed to
 gaseous) morphology studies in
 the higher redshift universe,  Block et
 al. (2001a) 
 present NASA-IRTF and SCUBA observations of NGC 922, a
 chaotic system in our local Universe which bears a striking
 resemblance to objects such as HDF 2-86 ($z=0.749$) in the HDF
 North.  If objects such as NGC 922 are common at
 high-redshifts, then this galaxy may serve as a local
 morphological `Rosetta stone' bridging low and high-redshift
 populations. Block et al. (2001a) show that quantitative
 measures of galactic structure are recoverable in the rest-frame
 infrared for NGC 922 seen at high redshifts using NGST,
 by simulating the appearance of this galaxy at redshifts z=0.7 and
 z=1.2 in rest-frame K$'$. 
{\em  Our results suggest that the capability of efficiently
 exploring
 the rest-wavelength IR morphology of high-z galaxies should probably be
 a key factor in deciding the final choice of instruments for the NGST}.

\section{Concluding Thoughts: Freeman and Ornithology}

The work of Ken Freeman in bars and galaxy dynamics may
be likened to his great passion for ornithology. Species such as
Aquila Rapax (the Tawny Eagle) and Aquila Nipalensis (the Steppe
Eagle) survey 
their terrain with exceptionally keen
foresight. Ken recognised
the fundamental nature of bars three decades ahead of the
commissioning of large format HgCdTe arrays in the early 1990s. 

\acknowledgments

DLB is indebted to Ken and to the SOC for their invitation to participate
in this Workshop. It is also a pleasure to thank Gary Da
Costa for his prompt reply to emails, to facilitate travel to Dunk Island.
RB would like to thank Ken for inspiring his early interest in
barred galaxies and resonance rings.
DLB is indebted to SASOL for sponsoring his sabbatical, during which
time this paper was prepared. In particular, he wishes to express his deep
appreciation to  Mr P. Kruger, Mr P. Cox and the entire SASOL Board. 
Without the financial support of the Anglo-American Chairman's Fund,
none of this research would have been possible. A special note of
thanks is expressed to the CEO of the Anglo-American Fund, Mrs M. Keeton.   
BGE was supported by grant AST-9870112 from the
U. S. National Science Foundation.


\begin{references}
\reference Block D.L. et al.
1994, \aap, 288, 365
\reference Block
D.L. \& Wainscoat R.J. 1996, Nature, 353, 48
\reference Block D.L.
\& Puerari I. 1999, \aap, 342, 627
\reference Block, D.L. et al. 2000,
 `Toward a New Millennium in Galaxy Morphology' (eds. D.L. Block,
 I. Puerari, A. Stockton \& D. Ferreira, Kluwer); see also \apss, 269, 5
\reference Block, D.L. et al. 2001a, \aap, 371, 393
\reference
Block, D.L. et al. 2001b, \aap (in press) [astro-ph/0106019]
 \reference Burstein D.
\& Rubin V. 1985, \apj, 297, 423
\reference
Buta, R.J. \& Combes, F. 1996, Fund. Cosmic Phys. 17, 95
\reference Buta, R.J. \& Block, D.L. 2001, \apj, 550, 243
\reference Combes, F. \&
Sanders, R. H. 1981, \aap, 96, 164
\reference
Curtis, H. D. 1918, Pub. Lick Obs. XIII, Part I, 11
\reference
de Vaucouleurs, G. 1959, Handbuch der Physik, 53, 275
\reference
de Vaucouleurs, G. \& Freeman, K.C. 1972, Vistas in Astronomy, 14, 163
\reference de Vaucouleurs, G., de Vaucouleurs, A, \& Freeman, K. C.
1968, \mnras, 139, 425
\reference Elmegreen, D.M. et al. 1999, \aj, 118, 2618
\reference
Eskridge, P. et al. 2000, AJ 119, 536
\reference Freeman, K.C. 1965, \mnras, 130, 183
\reference Freeman, K.C. 1966a, \mnras, 133, 47
\reference Freeman, K.C. 1966b, \mnras, 134, 1
\reference Freeman, K.C. 1966c, \mnras, 134, 15
\reference Freeman, K. C. 1992, in Physics
of Nearby Galaxies: Nature or Nurture?, T. X. Thuan, C. Balkowski, and
J. Tran Thanh Van, eds., Gif-sur-Yvette, Editions Frontiere, p. 201
\reference Grosb{\o}l, P.J. \& Patsis, P.A. 1998, \aap, 336, 840
\reference Herschel, J. 1947, `Results of astronomical observations
made during the years 1834-8 at the Cape of Good Hope', Smith and
Col., London.
\reference
 Hubble, E. 1926,
   \apj, 64, 321
\reference 
Knapen, J.H. 1999, in The Evolution of Galaxies on Cosmological
Timescales, J.E. Beckman \& T.J. Mahoney, Eds., ASP Conf. Ser.
187, 72
\reference
Knapen, J.H., Shlosman, I. \& Peletier, R.F. 2000, \apj, 529, 93
\reference Lindblad, P. A. B., Lindblad, P. O., \& Athanassoula, E.
1996, \aap, 313, 65
\reference Martin, P. 1995, \aj, 109, 2428
\reference Puerari, I. et al. 2000, \aap, 359, 932
\reference
Quillen, A. C., Frogel, J. A., \& Gonz\'alez, R. 1994, \apj, 437, 162
\reference
Sandage, A. \& Tammann, G. 1981, ``A Revised Shapley-Ames
Catalog
    of Bright Galaxies'' [RSA] (Carengie Inst; Wash. DC)
\reference Sanders, D.B. 2000, \apss, 269, 381
\reference
Sellwood, J.A. \& Wilkinson, A. 1993, Rep. Prog. Phys. 56,
173
\reference Zwicky, F. 1957, `Morphological Astronomy', Springer-Verlag, Berlin.
\end{references}
\end{document}